\documentclass[12pt]{article}
\pdfoutput = 1
\usepackage{graphics}
\usepackage[square,sort,comma,numbers,compress]{natbib}
\usepackage{array}
\usepackage{subfigure}
\usepackage{amsmath}
\usepackage{amssymb}
\usepackage{float}
\usepackage{graphicx} 
\usepackage[font=small,format=plain,labelfont=bf,up,textfont=normal,up]{caption}
\usepackage{multirow}
\usepackage{hhline}
\begin{document}
\date{Today}
\title{{\bf{\Large Viscosity to entropy density ratio for non-extremal Gauss-Bonnet black holes coupled to Born-Infeld electrodynamics }}}

\author{
{\bf {\normalsize Saurav Das}$^{a}$
\thanks{sd12ms101@iiserkol.ac.in, sauravdas110495@gmail.com}},\,
{\bf {\normalsize Sunandan Gangopadhyay}$^{a,b}
$\thanks{sunandan.gangopadhyay@gmail.com, sunandan@iiserkol.ac.in,  sunandan@associates.iucaa.in}},\,
{\bf {\normalsize Debabrata Ghorai}$^{c}$
\thanks{debanuphy123@gmail.com, debabrataghorai@bose.res.in}}\\
$^{a}$ {\normalsize Indian Institute of Science Education and Research Kolkata}\\{\normalsize Mohanpur, Nadia 741246, India}\\ 
$^{b}${\normalsize Visiting Associate in Inter University Centre for Astronomy \& Astrophysics,}\\
{\normalsize Pune 411007, India}\\
$^{c}${\normalsize  S.N. Bose National Centre for Basic Sciences,}\\{\normalsize JD Block, 
Sector III, Salt Lake, Kolkata 700106, India}\\[0.2cm]
}
\date{}

\maketitle

\begin{abstract}
\noindent The ratio of the shear viscosity to the entropy density is calculated for non-extremal Gauss-Bonnet (GB) black holes coupled to Born-Infeld (BI) electrodynamics in $5$ dimensions. The result is found to get corrections from the BI parameter and is analytically exact upto all orders in this parameter. The computations are then extended to D dimensions.
\end{abstract}
\vskip 1cm

\section{Introduction}
The investigation of transport properties of strongly coupled field theories using the tool of gauge/gravity correspondence \cite{adscft1} has been a prominent area of research in theoretical physics. One of the most interesting developments in this field has been the finding of the universal value $\frac{1}{4\pi}$ for the ratio of the shear viscosity to the entropy density. The dual gravity descriptions in which the computation was carried out have been a variety of theories described by Einstein gravity \cite{univ1}-\cite{univ10}, Gauss-Bonnet (GB) gravity \cite{gb1}-\cite{gb3}, GB gravity coupled to Maxwell electrodynamics \cite{tgc4} and in the presence of $F^4 $ corrections \cite{Kai}. The calculations were performed at a finite temperature for which the singularity structures of the black holes are different from those at zero temperature. These black holes with a non-zero Hawking temperature are known as non-extremal black holes. The result for the desired ratio has been shown to get both positive as well as negative corrections to the universal value of $\frac{1}{4\pi}$ for non-extremal black holes.

\noindent In case of extremal black holes for which the Hawking temperature vanishes, the ratio of the shear viscosity to the entropy density has been shown to be $\frac{1}{4\pi}$ from investigations carried out in the background of AdS RN black holes \cite{Kai2}, GB gravity with non-zero chemical potentials \cite{fd5} and more generally in \cite{fd12}.

\noindent In this paper, we calculate this ratio for field theories in the background of GB gravity coupled to BI electrodynamics at non-zero temperature. The importance of looking at this theory is that BI electrodynamics is the only non-linear theory of electrodynamics invariant under electromagnetic duality. It is also the most important non-linear electromagnetic theory free from infinite self energies of charged point particles that arises in the Maxwell theory \cite{bi}. The motivation for carrying out this investigation is to look for non-linear effects on the value of this ratio. 

\noindent This paper is organized as follows. In section 2, we provide the general set up to study the shear viscosity for non-extremal black hole which is coupled to Born-Infeld electrodynamics. In section 3, we compute the shear viscosity to entropy ratio for non-extremal black holes in five dimensions. In section 4, we compute the same in D dimensions. In section 5, we analytically calculate the shear viscosity to entropy ratio for GB gravity coupled to BI electrodynamics. We conclude finally in section 6.



\section{General set up}
In this section, we present the basic set up which would be needed to compute the shear viscosity in the background of non-extremal Gauss-Bonnet black hole in $5-$dimensions. The action for Gauss-Bonnet gravity with generic matter Lagrangian density is given by
\begin{eqnarray}
S=\frac{1}{16\pi G_5}\int d^5x \sqrt{-g}\left(R-2\Lambda+\frac{\lambda l^2}{2}(R_{\mu\nu\rho\sigma}R^{\mu\nu\rho\sigma}-4R_{\mu\nu}R^{\mu\nu}+R^2)+16\pi G_5 \mathcal{L}_M\right)
\label{eq1}
\end{eqnarray}
where $\lambda$ is the dimensionless Gauss-Bonnet parameter, $l$ is the radius of curvature of AdS space, $\mathcal{L}_M$ denotes the generic matter Lagrangian density and  $\Lambda=-\frac{6}{l^2}$ is the cosmological constant. For all future purposes in this work, we shall consider the matter Lagrangian density $\mathcal{L}_M$ to be electromagnetic in nature, that is
\begin{eqnarray}
\mathcal{L}_M=\mathcal{L}(F^2)
\label{eq2}
\end{eqnarray}
where $F^2=g^{\mu\rho}g^{\nu\sigma}F_{\mu\nu}F_{\rho\sigma}$ and $F_{\alpha\beta}=\partial_\alpha A_\beta - \partial_\beta A_\alpha$. \\
The equations of motions following from the above action admit solutions for the metric of the form
\begin{eqnarray}
ds^2=-H(r)N^2dt^2+\frac{dr^2}{H(r)}+\frac{r^2}{l^2}(dx^2+dy^2+dz^2)~.
\label{eq3}
\end{eqnarray}
In this paper, we restrict our discussions to those particular forms of $\mathcal{L}_{M}$ for which $H(r)$ can be written as
\begin{eqnarray}
H(r)=\frac{r^2}{l^2}G(r)
\label{eq4}
\end{eqnarray}
where $G(r)$ is regualar at $r=\infty $. Note that theories like matter-free Gauss-Bonnet gravity \cite{gb2}, Maxwell Gauss-Bonnet Gravity \cite{MGB} and the case in which we are interested, namely, Born-Infeld Gauss-Bonnet gravity \cite{BIM} satisfies the above condition. We now make a change of variable to simplify the subsequent calculations. This reads
\begin{eqnarray}
u=r_+^2/r^2
\label{eq5}
\end{eqnarray}
where $r_+$ denotes the radius of the outer horizon. The metric in this new coordinates takes the form
\begin{eqnarray}
ds^2=\frac{1}{l^2a^2u}\left(-f(u)N^2dt^2+d\vec{x}^2\right)+\frac{l^2}{4u^2f(u)}du^2
\label{eq6}
\end{eqnarray}
where $d\vec{x}^2=(dx^2+dy^2+dz^2)$, $a=1/r_+$ and $f(u)=G(r)$. Thermodynamic quantities in the new coordinates are given by
\begin{eqnarray}
T &=& -\frac{Nr_+}{2\pi l^2}f'(1) \\
s &=& \frac{1}{4G_5}\frac{r_+^3}{l^3}
\label{eq7}
\end{eqnarray}
where $\prime$ denotes derivative with respect to $u$.

\section{Shear viscosity to entropy ratio for non-extremal black holes}
Kubo's formalism involving retarded Green's functions give a way to compute the shear viscosity of the conformal field theory living on the boundary of the AdS black holes. The shear viscosity is given by
\begin{eqnarray}
\eta = \lim_{\omega\rightarrow 0} - \frac{Im[ G^{R}_{xy,xy} (\omega, 0)]}{\omega}~.
\label{eta}
\end{eqnarray} 
The retarded Green's function of the energy-momentum tensor of the boundary field theory dual to the bulk theory is calculated by making a small tensor perturbation of the background metric denoted by
\begin{eqnarray}
\phi(t,z,u)=h^{~y}_x(t,z,u)~.
\label{eq8}
\end{eqnarray}
The perturbed metric therefore takes the form
\begin{eqnarray}
ds^2=\frac{1}{l^2a^2u}\left(-f(u)N^2dt^2+d\vec{x}^2+2\phi(t,z,u)dxdy\right)+\frac{l^2}{4u^2f(u)}du^2  ~.
\label{eq9}
\end{eqnarray}
We now introduce the Fourier transform of $\phi(t,z,u)$ :  
\begin{eqnarray}
\phi(t,z,u)=\int\frac{d^4k}{(2\pi)^4}e^{-i\omega t+ik_3z}\phi(k,u)
\label{eq10}
\end{eqnarray}
where $k=(\omega, 0,0,k_3)$ and $\phi (-k, u) = \phi^{*} (k, u)$. Plugging this form into the action (\ref{eq1}) and using eq.(\ref{eq9}), we compute the effective graviton action upto second order in $\phi(k,u)$. This reads
\begin{eqnarray}
S_{eff}[\phi(k,u)]=-\frac{r_+^4N}{16\pi G_5l^5}\int\frac{d^4k}{(2\pi)^4}du\left[g(u)\phi'^2(u)-g_2(u)\phi^2(u) \right] 
\label{eq11}
\end{eqnarray}
upto terms involving total derivatives. Note that
\begin{eqnarray}
\label{eq12}
g(u) &=& \frac{f(u)}{u}\left[1+2\lambda u^2\left(\frac{f(u)}{u}\right)'\right]\\
g_2(u) &=& g(u)\frac{\overline{\omega}^2}{uN^2f^2(u)}-\overline{k}_3^2u^{-2}\left[1-2\lambda u^2\left\{ 2u\left(\frac{f(u)}{u}\right)''+3\left(\frac{f(u)}{u}\right)'\right\} \right]
\label{eq13}
\end{eqnarray}
where $\overline{\omega}=\frac{l^2a}{2}\omega$ and $\overline{k}_3=\frac{l^2a}{2}k_3$.\\
Since the calculation of shear viscosity involves taking the zero momentum limit, we can set $k_3=0$. The equation of motion in $\phi(k,u)$ then reads
\begin{eqnarray}
\phi''(u)+\frac{g'(u)}{g(u)}\phi'(u)+\frac{\overline{\omega}^2}{uN^2f^2(u)}\phi(u)=0~.
\label{eq14}
\end{eqnarray}
We now proceed to solve this equation of motion for the case of the non-extremal black hole for which $T \neq 0 $ and hence $f^{\prime}(1) \neq 0$ by assuming the solution to be of the form \cite{MGB}
\begin{eqnarray}
\phi(u)=(1-u)^{\nu}F(u)
\label{eqA1}
\end{eqnarray}
where $F(u)$ is regular at $u=1$. Putting this in eq.(\ref{eq14}) and looking at the asymptotic behaviour of this equation near $u=1$ yields the value of $\nu$ to be
\begin{eqnarray}
\nu=\pm \frac{i\omega}{4\pi T}~.
\label{eqA2}
\end{eqnarray}
Since we are interested in the incoming wave, hence we consider only the negative root $\nu = -\frac{i\omega}{4\pi T}$. 
Substituting eq.(\ref{eqA1}) in eq.(\ref{eq14}) and keeping terms upto linear order in $\nu$, we get
\begin{eqnarray}
\left(g(u)F'(u)\right)' -\frac{2\nu}{1-u}g(u)F^{\prime}(u) - \nu\left(\frac{g(u)}{1-u}\right)F(u)=0~~.
\label{eqA4}
\end{eqnarray} 
We now expand around $\omega=0$ since we require only the low frequency behaviour of the solution to calculate the shear viscosity
\begin{eqnarray}
F(u)=F^{(0)}(u)+\nu F^{(1)}(u)+\cdots
\label{eqA3}
\end{eqnarray}
Eq.(\ref{eqA4}) is solved recursively by substituting eq.(\ref{eqA3}) in eq.(\ref{eqA4}) to find $F^{(0)}(u)$ and $F^{(1)}(u)$. Regularity condition of $F(u)$ \cite{tgc4} leads to
\begin{eqnarray}
F^{(0)}(u) &=& C =constant
\label{eqA5}
\end{eqnarray}
\begin{eqnarray}
F^{(1)\prime}(u) &=& \frac{C}{1-u}+\frac{Cf'(1)\left[1+2\lambda f'(1)\right]}{g(u)}
\label{eq17}
\end{eqnarray}
which can be integrated to yield
\begin{eqnarray}
F^{(1)}(u)=-C\ln(1-u)+Cf'(1)\left[1+2\lambda f'(1)\right]\int\frac{du}{g(u)}+D
\label{eqnew1}
\end{eqnarray}
The constant $C$ can be determined in terms of the value of the field $\phi (u)$ at the boundary $u=0$ which implies
\begin{eqnarray}
\label{eq18}
\lim_{u\rightarrow 0} \phi (u) &=& \phi^{(0)}(u) \nonumber \\
\Rightarrow C &=& \phi^{(0)}(u)\left[ 1+O(\nu) \right]~.
\end{eqnarray}
The on-shell graviton action is now computed by using the equation of motion (\ref{eq14}) in the effective graviton action (\ref{eq3}). This yields
\begin{eqnarray}
S_{on-shell}[\phi(k,u)]=-\frac{r_+^4N}{16\pi G_5l^5}\int\frac{d^4k}{(2\pi)^4}\left[ g(u)\phi (u) \phi'(u)\right]\Bigg\rvert_{u=0} ~.
\label{eq19}
\end{eqnarray}
Substituting the expression for $\phi(u)$ from eq.(\ref{eqA1}) in eq.(\ref{eq19}) and using eq.(s)(\ref{eqA2},\ref{eq17}), the on-shell action (\ref{eq19}) takes the form
\begin{eqnarray}
S_{on-shell}=\int\frac{d^4k}{(2\pi)^4}\phi^{(0)}(k)\mathcal{G}_{xy,xy}(k,u)\phi^{(0)}(-k)\Bigg\rvert_{u=0}
\label{on}
\end{eqnarray}
where 
\begin{eqnarray}
\mathcal{G}_{xy,xy}(k,u)=\frac{-i\omega}{2} \frac{1}{16 \pi G_5}\left(\frac{r_+}{l}\right)^3\left[1+2\lambda f'(1)\right]+O(\omega^2). 
\label{om}
\end{eqnarray}
The retarded Green's function can be calculated from this on-shell action from the expression \cite{NB}
\begin{eqnarray}
G^R_{xy,xy}=\lim_{u\rightarrow 0}2\mathcal{G}_{xy,xy}(k,u)~.
\label{eqnew4}
\end{eqnarray}
From this we obtain the expression for the retarded Green's function to be
\begin{eqnarray}
G^R_{xy,xy}(\omega,0)= -i\omega \frac{1}{16 \pi G_5}\left(\frac{r_+}{l}\right)^3\left[1+2\lambda f'(1)\right]+O(\omega^2)~.
\label{eq20}
\end{eqnarray}
The shear viscosity can now be calculated using eq.(\ref{eta}) and reads
\begin{eqnarray}
\eta= \frac{1}{16 \pi G_5}\left(\frac{r_+}{l}\right)^3\left[1+2\lambda f'(1)\right].
\label{eq22}
\end{eqnarray} 
Hence the ratio of the shear viscosity to entropy is given by
\begin{eqnarray}
\frac{\eta}{s}=\frac{1}{4\pi}\left[1+2\lambda f'(1)\right]~.
\label{eq23}
\end{eqnarray} 
This is the most general form for the $\frac{\eta}{s}$ ratio for non-extremal black holes.
It is assuring to note that our expression matches with specific cases for matter free Gauss-Bonnet black hole \cite{gb2} and Gauss-Bonnet black hole with Maxwell electrodynamics \cite{MGB}.\\
For the AdS Gauss Bonnet black hole in five dimensions
\begin{eqnarray}
f(u)=\frac{1}{2\lambda}\left[1-\sqrt{1-4\lambda(1-u^2)}\right]
\label{eqnew5}
\end{eqnarray}
which yields
\begin{eqnarray}
f'(1)=-2
\label{eqnew6}
\end{eqnarray}
Putting this value for $f'(1)$ in eq.(\ref{eq23}) gives \cite{gb3}
\begin{eqnarray}
\frac{\eta}{s}=\frac{1}{4\pi}(1-4\lambda).
\label{eqnew7}
\end{eqnarray}
For the AdS GB black hole in Maxwell electrodynamics
\begin{eqnarray}
f(u)=\frac{1}{2\lambda}\left[1-\sqrt{1-4\lambda(1-u)(1+a-au^2)}\right]
\label{eqnew8}
\end{eqnarray}
where $a=\frac{q^2l^2}{r_+^6}$.
This gives
\begin{eqnarray}
f'(1)=-(2-a)
\label{eqnew9}
\end{eqnarray}
which finally leads to \cite{MGB}
\begin{eqnarray}
\frac{\eta}{s}=\frac{1}{4\pi}\left[1-4\lambda\left(1-\frac{a}{2}\right)\right].
\label{eqnew10}
\end{eqnarray}

\section{Non-extremal black holes in D dimensions}
In this section, we proceed to generalize the result in eq.(\ref{eq23}) in D dimensions. The action for Gauss-Bonnet gravity in D dimensions can be written as
\begin{equation}
\begin{split}
& S=\frac{1}{16\pi G_D}\int d^Dx \sqrt{-g}\times \\
& \left(R-2\Lambda+\frac{\lambda l^2}{(D-3)(D-4)}(R_{\mu\nu\rho\sigma}R^{\mu\nu\rho\sigma}-4R_{\mu\nu}R^{\mu\nu}+R^2)+16\pi G_D \mathcal{L}_M\right)
\label{eqD1}
\end{split}
\end{equation}
where $\Lambda=-\frac{(D-1)(D-2)}{2l^2}$. Following the analysis in the previous section, we obtain the effective action in $\phi(k,u)$ to be 
\begin{eqnarray}
S_{eff}[\phi(k,u)]=-\frac{r_+^{D-1}N}{16\pi G_Dl^D}\int\frac{d^4k}{(2\pi)^4}du\left[v(u)\phi'^2(u)-v_2(u)\phi^2(u)\right]
\label{eqD2}
\end{eqnarray}
up to total derivative terms. The functions $v(u)$ and $v_2(u)$ in this case are given by
\begin{eqnarray}
v(u)=u^{-\frac{D-3}{2}}f(u)\left[1-\frac{2\lambda}{D-3}\left(-2u^2\left[\frac{f(u)}{u}\right]'+(D-5)f(u)\right)\right]
\label{eqD3}
\end{eqnarray}
\begin{equation}
\begin{split}
& v_2(u)=v(u)\frac{\overline{\omega}^2}{uN^2f^2(u)}-\overline{{k}_3^2}u^{-\frac{D-1}{2}}\times \\
& \left[1-\frac{2\lambda}{(D-3)(D-4)}\left(2u^2\left[2u\left(u^{-1}f(u)\right)''+3\left(u^{-1}f(u)\right)'\right] \right. \right. \\
& \left. \left. -(D-5)\left[4u^2\left(u^{-1}f(u)\right)'-(D-6)f(u)\right]\right)\right].
\end{split}
\label{eqD4}
\end{equation}
The equation of motion for $\phi(u)$ in the zero momentum limit reads
\begin{eqnarray}
\phi''(u)+\frac{v'(u)}{v(u)}\phi'(u)+\frac{v_2(u)}{v(u)}\phi(u)=0.
\label{eqD5}
\end{eqnarray}
Once again assuming the solution of the above equation to be of the form in eq.(\ref{eqA1}) yields
\begin{eqnarray}
F^{(1)\prime}(u) &=& \frac{C}{1-u}+\frac{Cf'(1)\left[1+\frac{4\lambda}{D-3} f'(1)\right]}{v(u)}~.
\label{eqD6}
\end{eqnarray}
After carrying out the analysis as in the previous section with the above result, we arrive at
\begin{eqnarray}
\frac{\eta}{s}=\frac{1}{4\pi}\left[1+\frac{4\lambda}{D-3}f'(1)\right]~.
\label{eqD7}
\end{eqnarray} 
The above result reduces to eq.(\ref{eq23}) for $D=4$.
\section{Gauss Bonnet black hole in Born-Infeld electrodynamics}
\begin{figure}[H]
  \centering
  \begin{minipage}[b]{0.4\textwidth}
    \includegraphics[height=15em]{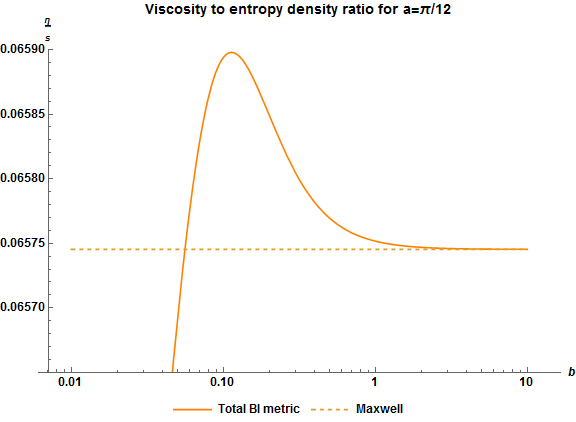}
    \caption{Viscosity to entropy ratio for small value of BI parameter}
		\label{fig:small_b}
  \end{minipage}
  \hfill
  \begin{minipage}[b]{0.4\textwidth}
    \includegraphics[height=15em]{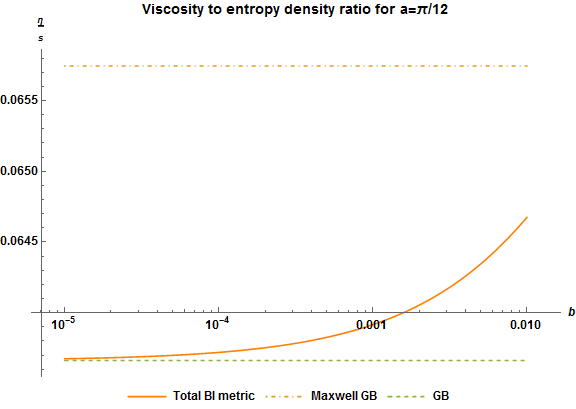}
    \caption{Viscosity to entropy ratio in the limit of zero charge}
		\label{fig:b0}
  \end{minipage}
\end{figure}

\begin{figure}[H]
  \centering
  \begin{minipage}[b]{0.4\textwidth}
    \includegraphics[height=15em]{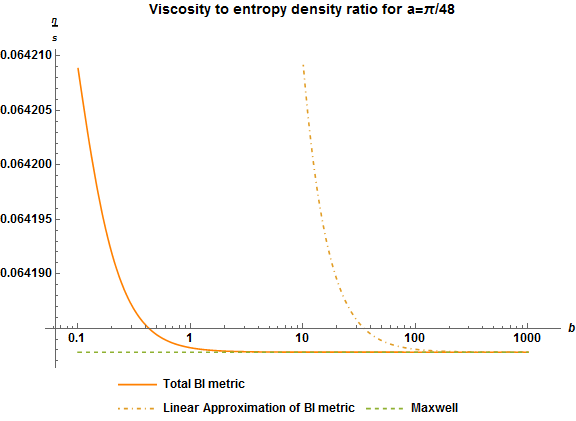}
    \caption{Viscosity to entropy ratio for small charge regime}
		\label{fig:small_a}
  \end{minipage}
  \hfill
  \begin{minipage}[b]{0.4\textwidth}
    \includegraphics[height=15em]{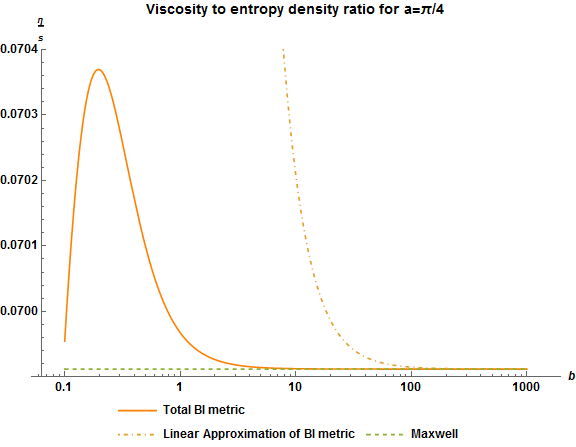}
    \caption{Viscosity to entropy ratio in large charge regime}
		\label{fig:a1}
  \end{minipage}
\end{figure}

In this section, we calculate the $\frac{\eta}{s}$ ratio for the non-extremal Gauss-Bonnet black hole in the presence of Born-Infeld electrodynamics. The Lagrangian density for Born-Infeld electrodynamics reads
\begin{eqnarray}
\mathcal{L}_M=\mathcal{L}_{BI}(F^2)=4b^2\left(1-\sqrt{1+\frac{F^2}{2b^2}}\right)
\label{eq24}
\end{eqnarray}
where $b$ is the Born-Infeld parameter. In the limit $b\rightarrow \infty$, one recovers Maxwell electrodynamics. Assuming a solution of the equations of motion in the form (\ref{eq3}), H(r) is given by \cite{MGB}
\begin{eqnarray}
H(r) &=& \frac{r^2}{2\lambda l^2}\times \nonumber \\
&& \left[1- \sqrt{1-4\lambda\left\{1-\frac{\mu l^2}{r^4}+\frac{16}{3}\pi Gb^2l^2\left(1-\sqrt{1+\frac{q^2}{b^2r^6}}\right)+8\pi Gl^2\frac{q^2}{r^6} {}_2F_1\left[\frac{1}{3},\frac{1}{2},\frac{4}{3},-\frac{q^2}{b^2r^6}\right]\right\}}\right] \nonumber \\
\label{eq25}
\end{eqnarray}
where $F_1[.,.,.,.]$ denotes the Gauss Hypergeometric function, $\mu$ is an integration constant and q is another integration constant related to charge. \\
\noindent Now $N^2$ can be determined from the condition
\begin{eqnarray}
\lim_{r\rightarrow \infty}H(r)N^2 &=& \frac{r^2}{l^2}
\label{eq26}
\end{eqnarray}
which implies
\begin{eqnarray}
 N^2 &=& \frac{1}{2}\left(1+\sqrt{1-4\lambda}\right)~.
\end{eqnarray}
Defining new parameters
\begin{eqnarray}
a_1=\frac{\mu l^2}{r_+^4} \ \ \ \ \ \ a_2=\frac{16\pi Gl^2q^2}{3r_+^6} \ \ \ \ \ \ c_1=\frac{16}{3}\pi Gl^2
\label{eq27}
\end{eqnarray}
the function $f(u)$ takes the form
\begin{eqnarray}
f(u)&=&\frac{1}{2\lambda} \times \nonumber \\ 
&& \left[1- \sqrt{1-4\lambda\left(1-a_1u^2+c_1b^2\left(1-\sqrt{1+\frac{a_2u^3}{c_1b^2}}\right)+\frac{3}{2}a_2u^3{}_2F_1\left[\frac{1}{3},\frac{1}{2},\frac{4}{3},-\frac{a_2u^3}{c_1b^2}\right]\right)}\right]. \nonumber \\
\label{eq28}
\end{eqnarray}
Since $f(1) = 0 $, the parameters $a_1$ and $a_2$ are related by
\begin{eqnarray}
a_1-c_1b^2+c_1b^2\sqrt{1+\frac{a_2}{c_1b^2}}-\frac{3a_2}{2}{}_2F_1\left[\frac{1}{3},\frac{1}{2},\frac{4}{3},-\frac{a_2}{c_1b^2}\right]=1~.
\label{eq29}
\end{eqnarray}
Using this relation in eq.(\ref{eq28}), we obtain
\begin{eqnarray}
f'(1)=-\left[2a_1+\frac{3a_2}{2\sqrt{1+\frac{a_2}{c_1b^2}}}-\frac{9a_2}{2}{}_2F_1\left[\frac{1}{3},\frac{1}{2},\frac{4}{3},-\frac{a_2}{c_1b^2}\right]+\frac{9a_2^2}{36c_1b^2}{}_2F_1\left[\frac{4}{3},\frac{3}{2},\frac{7}{3},-\frac{a_2}{c_1b^2}\right]\right].~
\label{eq30}
\end{eqnarray}
The $\frac{\eta}{s}$ ratio is now fully determined from Eq.(s)(\ref{eq23}) and (\ref{eq30}). The above equation indicates that the viscosity to entropy density ratio of a five dimensional AdS-GB black hole in BI electrodynamics is always less than $\frac{1}{4\pi}$. The ratio approaches the standard Maxwell GB value \cite{MGB} in the $b\rightarrow \infty$ limit. This is evident from the Figure $[\ref{fig:small_a}]$ and $[\ref{fig:a1}]$). Figure $[\ref{fig:b0}]$ shows that it approaches the free GB value \cite{gb2} in the $b\rightarrow 0$ limit. Figure $[\ref{fig:small_b}]$ shows the viscosity to entropy density ratio for AdS GB black hole coupled to BI ED is smaller than that for  AdS GB black hole coupled to Maxwell ED for small value of BI parameter, then there is a switch over and finally for high values of the BI parameter, the ratios become identical.
\noindent We now simplify our results by looking at the leading order contribution of the Born-Infeld parameter b in the metric $H(r)$. For this we expand $H(r)$ and take only the leading order term in $\frac{1}{b}$. This yields
\begin{eqnarray}
H(r)=\frac{r^2}{2\lambda l^2}\left(1-\sqrt{1-4\lambda-\frac{64\pi Gl^2\lambda q^2}{3r^6}+\frac{4\mu l^2\lambda}{r^4}}\right)-\frac{G\pi q^4}{\left(3r^{10}\sqrt{1-4\lambda-\frac{64\pi Gl^2\lambda q^2}{3r^6}+\frac{4\mu l^2\lambda}{r^4}}\right)b^2}. \nonumber \\
\label{eq31}
\end{eqnarray}
Defining a new parameter
\begin{eqnarray}
c_2=\frac{3}{128\pi Gl^2}
\label{eq32}
\end{eqnarray}
the functional form of $f(u)$ upto order $\frac{1}{b^2}$ takes the form
\begin{eqnarray}
f(u)=\frac{1}{2\lambda}\left[1-\sqrt{1-4\lambda\left(1-a_1u^2+a_2u^3\right)}-\frac{c_2\lambda a_2^2u^6}{b^2\sqrt{1-4\lambda\left(1-a_1u^2+a_2u^3\right)}}\right]~.
\label{eq33}
\end{eqnarray}
As $f(1)=0$, the constants $a_1$ and $a_2$ are related by
\begin{eqnarray}
\label{eq34}
\sqrt{1-4\lambda\left(1-a_1+a_2\right)}+\frac{a_2^2c_2}{b^2\sqrt{1-4\lambda\left(1-a_1+a_2\right)}}=1~.
\end{eqnarray}
Eq(s).(\ref{eq33}) and (\ref{eq34}) yield
\begin{eqnarray}
f'(1)=\frac{3a_2-2a_1}{\sqrt{1-4\lambda\left(1-a_1+a_2\right)}}+\frac{a_2^2c_2}{b^2}\left(\frac{12\lambda(1-a_1+a_2)-[3+\lambda(3a_2-2a_1)]}{\left(1-4\lambda\left(1-a_1+a_2\right)\right)^3}\right)~.
\label{eq35}
\end{eqnarray}
It is reassuring to note that the exact result (\ref{eq30}) when expanded upto leading order in $\frac{1}{b}$ yields eq.(\ref{eq35}).

\noindent We now move on to calculate the ratio $\frac{\eta}{s}$ for the D dimensional AdS-GB black hole coupled to BI electrodynamics.\\
The black hole metric in D dimensions reads,
\begin{equation}
\begin{split}
& H(r)=\frac{r^2}{2\lambda l^2} \times \\
& \left(1-\left[1-4\lambda\left(1-\frac{\mu l^2}{r^{D-1}}+\frac{64\pi Gl^2b^2}{(D-1)(D-2)}\left(1-\sqrt{1+\frac{q^2}{b^2r^{2D-4}}}\right) \right. \right. \right. \\
& \left. \left. \left. +\frac{64\pi Gl^2q^2}{(D-1)(D-3)r^{2D-4}}{}_2F_1\left[\frac{1}{2},\frac{D-3}{2D-4},\frac{3D-7}{2D-4},-\frac{q^2}{b^2r^{2D-4}}\right]\right)\right]^{\frac{1}{2}}\right)~.
\label{eqD11}
\end{split}
\end{equation} 
Defining
\begin{eqnarray}
a_{1D}=\frac{\mu l^2}{r_+^{D-1}} \ \ \ \ \ \ \  a_{2D}=\frac{16\pi Gl^2q^2}{3r_+^{2D-4}} \ \ \ \ \ \ \ c_1=\frac{16}{3}\pi Gl^2
\label{eqD12}
\end{eqnarray}
we have
\begin{equation}
\begin{split}
f(u)=\frac{1}{2\lambda} \times \\
&\quad \left(1-\left[1-4\lambda\left(1-a_{1D}u^{\frac{D-1}{2}}+\frac{12c_1b^2}{(D-1)(D-2)}\left(1-\sqrt{1+\frac{a_{2D}u^{D-2}}{c_1b^2}}\right)\right. \right. \right. \\
& \left. \left. \left. +\frac{12a_{2D}u^{D-2}}{(D-1)(D-3)}{}_2F_1\left[\frac{1}{2},\frac{D-3}{2D-4},\frac{3D-7}{2D-4},-\frac{a_{2D}u^{D-2}}{c_1b^2}\right]\right)\right]^{\frac{1}{2}}\right)~.
\label{eqD13}
\end{split}
\end{equation}
Since $f(1)=0$, the constants $a_{1D}$ and $a_{2D}$ are not independent and are related by
\begin{eqnarray}
a_{1D}=1+\frac{12c_1b^2}{(D-1)(D-2)}\left(1-\sqrt{1+\frac{a_{2D}}{c_1b^2}}\right)+\frac{12a_{2D}}{(D-1)(D-3)}{}_2F_1\left[\frac{1}{2},\frac{D-3}{2D-4},\frac{3D-7}{2D-4},-\frac{a_{2D}}{c_1b^2}\right]~.
\label{eqD14}
\end{eqnarray}
Eqs. (\ref{eqD13}) and (\ref{eqD14}) finally gives
\begin{equation}
\begin{split}
& f'(1)=-\frac{(D-1)a_{1D}}{2}-\frac{6a_{2D}}{(D-1)}{}_2F_1\left[\frac{1}{2},\frac{D-3}{2D-4},\frac{3D-7}{2D-4},-\frac{a_{2D}}{c_1b^2}\right]\\
& +\frac{12a_{2D}(D-2)}{(D-3)(D-1)}{}_2F_1\left[\frac{1}{2},\frac{D-3}{2D-4},\frac{3D-7}{2D-4},-\frac{a_{2D}}{c_1b^2}\right]~.
\label{eqD15}
\end{split}
\end{equation}
The viscosity to entropy density ratio for AdS-GB black hole coupled to D dimensional space-time is completely specified by eq(s).(\ref{eqD7}) and (\ref{eqD15}).\\
Fig. $5$ shows that the viscosity to entropy density ratio for a fixed value of the extremality parameter (that is $\frac{q^2}{r_+^{2D-4}}$) increases with dimension of the spacetime, but always stays below the extremal limit of $\frac{1}{4\pi}$.
\begin{figure}[H]
	\centering
		\includegraphics[height=15em]{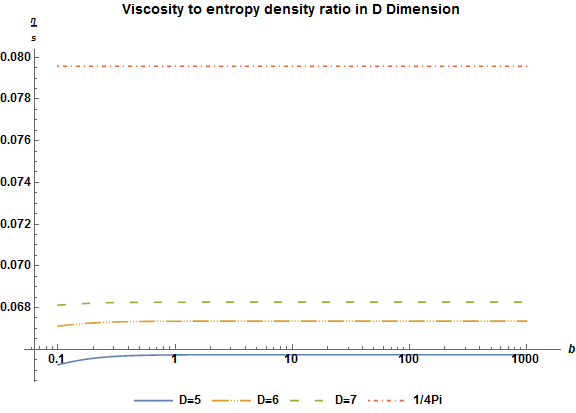}
	\caption{Figure representing viscosity to entropy density ratio for different dimensions}
	\label{fig:D-Dimensions}
\end{figure}

\section{Conclusions}
In this paper, we compute the shear viscosity to entropy density ratio for field theories in the background of GB black holes (at non-zero temperature) coupled to BI electrodynamics. We find that the ratio gets corrections due to the BI coupling parameter.\\
We observe that the $\frac{\eta}{s}$ ratio is always less than $\frac{1}{4\pi}$. The result reduces to the standard Maxwell GB value in the $b\rightarrow\infty$ limit and to the GB value in the $b\rightarrow 0$ limit. We then extend our computations in case of the D-dimensional AdS GB black hole coupled to BI electrodynamics. Here we also find that the $\frac{\eta}{s}$ ratio gets corrected due to the BI parameter and also depends upon the dimension D of the black hole spacetime. In particular we observe that the $\frac{\eta}{s}$ ratio for a fixed value of the extremality parameter increases with increase in the dimension of the black hole spacetime, but always remain lower than the value $\frac{1}{4\pi}$.

\section*{Acknowledgments} S.G. acknowledges the support by DST SERB under Start Up Research Grant (Young Scientist), File No.YSS/2014/000180. DG would like to thank DST-INSPIRE, Govt. of India for financial support.



\begin{thebibliography}{99}
\baselineskip=0.6 cm
\bibitem{adscft1} J. M. Maldacena, Adv. Theor. Math. Phys. 2, 231 (1998).
\bibitem{univ1} G. Policastro, D.T. Son and A.O. Starinets, Phys. Rev. Lett. 87 (2001) 081601.
\bibitem{univ2} G. Policastro, D.T. Son and A.O. Starinets, JHEP 09 (2002) 043.
\bibitem{univ3} G. Policastro, D.T. Son and A.O. Starinets, JHEP 12 (2002) 054.
\bibitem{univ4} A. Buchel and J.T. Liu, Phys. Rev. Lett. 93 (2004) 090602.
\bibitem{univ5} J. Mas, JHEP 03 (2006) 016.
\bibitem{univ6} D.T. Son and A.O. Starinets, JHEP 03 (2006) 052.
\bibitem{univ7} O. Saremi, JHEP 10 (2006) 083.
\bibitem{univ8} K. Maeda, M. Natsuume and T. Okamura, Phys. Rev. D 73 (2006) 066013.
\bibitem{univ9} R.G. Cai and Y.-W. Sun, JHEP 09 (2008) 115.
\bibitem{univ10} H.S. Tan, JHEP 04 (2009) 131.
\bibitem{gb1} Y. Kats and P. Petrov, JHEP 01 (2009) 044.
\bibitem{gb2} M. Brigante, H. Liu, R. C. Myers, S. Shenker, S. Yaida, Phys. Rev. D 77, 126006 (2008).
\bibitem{gb3} M. Brigante, H. Liu, R.C. Myers, S. Shenker, S. Yaida,  Phys. Rev. Lett. 100 (2008) 191601.
\bibitem{tgc4} X.-H. Ge, Y. Matsuo, F.-W. Shu, S.-J. Sin and T. Tsukioka, JHEP 10 (2008) 009.
\bibitem{Kai} R.G. Cai, Z.Y. Nie, and Y.W. Sun, Phys. Rev. D 78 126007 (2008)
\bibitem{Kai2} R. G. Cai, Yan Liu, Ya-Wen Sun, JHEP 04 (2010)090.
\bibitem{fd5} R.G. Cai, Z.-Y. Nie, N. Ohta and Y.-W. Sun, Phys. Rev. D 79 (2009) 066004.
\bibitem{fd12} S.S. Pal, Phys. Rev. D 81 (2010) 045005.


\bibitem{bi}M.~Born, L.~Infeld, Foundations of the new field theory, Proc. Roy. Soc. Lond. A 144 (1934) 425.
\bibitem{MGB} X. H. Ge, Y. Matsuo, F. W. Shu, S. J. Sin, T. Tsukioka, JHEP 10 (2008)009.
\bibitem{BIM} O. Miskovic, R. Olea, Phys. Rev. D 83, 024011 (2011).
\bibitem{NB} N. Bannerjee, ``Different Aspects of Black Hole Physics in String Theory", May,2009.

\end{thebibliography}
\end{document}